\documentclass[aps,pre,superscriptaddress,preprint,
amsmath,amssymb,floatfix]{revtex4-1}
\usepackage{graphicx}
\usepackage{bm}
\usepackage[colorlinks]{hyperref}
\usepackage{siunitx}

\newcommand{\AVE}[1]{\ensuremath{\left\langle {#1} \right\rangle}}
\newcommand{\ABS}[1]{\ensuremath{\left\lvert {#1} \right\vert}}
\newcommand{\dpone}[2]{\ensuremath{\displaystyle\frac{\partial {#1}}{\partial
			{#2}}}}
\newcommand{\dptwo}[2]{\ensuremath{\displaystyle\frac{\partial^2 {#1}}{\partial
			{#2}^2}}}

\newcommand{\bB}{\ensuremath{\bm{B}}}

\newcommand{\bD}{\ensuremath{\bm{D}}}

\newcommand{\bF}{\ensuremath{\bm{F}}}

\newcommand{\bH}{\ensuremath{\bm{H}}}
\newcommand{\bI}{\ensuremath{\bm{I}}}

\newcommand{\bL}{\ensuremath{\bm{L}}}

\newcommand{\bQ}{\ensuremath{\bm{Q}}}

\newcommand{\bU}{\ensuremath{\bm{U}}}

\newcommand{\be}{\ensuremath{\bm{e}}}

\newcommand{\bj}{\ensuremath{\bm{j}}}

\newcommand{\bmm}{\ensuremath{\bm{m}}}
\newcommand{\bn}{\ensuremath{\bm{n}}}

\newcommand{\bq}{\ensuremath{\bm{q}}}

\newcommand{\bt}{\ensuremath{\bm{t}}}
\newcommand{\bu}{\ensuremath{\bm{u}}}

\newcommand{\bx}{\ensuremath{\bm{x}}}

\newcommand{\bz}{\ensuremath{\bm{z}}}

\newcommand{\bsigma}{\ensuremath{\bm{\sigma}}}

\newcommand{\bomega}{\ensuremath{\bm{\omega}}}

\newcommand{\hsigma}{\ensuremath{\hat{\sigma}}}

\DeclareMathOperator{\Tr}{Tr}

\newcommand{\ERF}{\mathrm{Erf}}
\newcommand{\ERFI}{\mathrm{Erfi}}

\begin{document}

\title{Anisotropic swim stress in active matter with nematic order}
	\author{Wen Yan}
	\email[]{wyan@flatironinstitute.org}
	\affiliation{Department of Mechanical \& Civil Engineering, 
		Division of Engineering \& Applied Science, 
		California Institute of Technology, Pasadena, CA 91125}
	\altaffiliation{Current Address 1: Center for Computational Biology, Flatiron Institute, Simons
		Foundation, New York, NY 10010}
	\altaffiliation{Current Address 2: Courant Institute of Mathematical Sciences, New York University, New York, NY 10012}
	\author{John F. Brady}
	\email[]{jfbrady@caltech.edu}
	\affiliation{Division of Chemistry \& Chemical Engineering 
		and Division of Engineering \& Applied Science, 
		California Institute of Technology, Pasadena, CA 91125}

	\date{\today}

\begin{abstract}
Active Brownian Particles (ABPs)  transmit a swim pressure $\Pi^{swim}=n\zeta D^{swim}$ to the container boundaries, where $\zeta$ is the drag coefficient, $D^{swim}$ is the swim diffusivity and $n$ is the uniform bulk number density far from the container walls. In this work we extend the notion of the isotropic swim pressure to the anisotropic tensorial swim stress $\bsigma^{swim} = - n \zeta \bD^{swim}$, which is related to the anisotropic swim diffusivity $\bD^{swim}$. We demonstrate this relationship with ABPs that achieve nematic orientational order via a bulk external field. The anisotropic swim stress is obtained analytically for dilute ABPs in both 2D and 3D systems, and the anisotropy is shown to grow exponentially  with the strength of the external field. We verify that the normal component of the anisotropic swim stress applies a pressure $\Pi^{swim}=-(\bsigma^{swim}\cdot\bn)\cdot\bn$ on a wall with normal vector $\bn$, and, through Brownian dynamics simulations, this pressure is shown  to be the force per unit area transmitted by the active particles.  Since ABPs have no friction with a wall, the difference between the normal and tangential stress components -- the normal stress difference --  generates a net flow of ABPs along the wall, which is a generic property of active matter systems.
\end{abstract} 
 
\maketitle


\textbf{Introduction.} In active matter each  particle propels itself with a velocity $U_0$ along a direction characterized by an orientation vector $\bq$, and  by manipulating $\bq$, either as a result of some intrinsic reorientation mechanism (e.g. Brownian torques) or in response to an external field,  interesting phenomena arise, such as shear trapping \cite{Rusconi2014}, rheotaxis \cite{Kaya2012}, action-at-distance \cite{SwimForce2015}, etc. These phenomena can be captured via particle-level Langevin dynamic simulation of the simple Active Brownian Particles (ABPs) model, or by solving the corresponding Smoluchowski equation for the probability density in position and orientation space. 

As a complement to the Smoluchowski analysis, continuum mechanics may also be applicable and provide a simpler description in the large-scale  to determine the deformation and flow of active matter.  The detailed dynamics at the Smoluchowski level are encapsulated into the balance of forces and stresses at the continuum scale---a balance of body and surface forces.
The surface force of active matter is the swim pressure \cite{Pressure2014}, which is the pressure required to confine the swimmers within a volume, and, like the osmotic pressure of passive Brownian particles, the swim pressure is related to the swim diffusivity: $\bsigma^{swim}=-n\zeta D^{swim}\bI$, 
where $n$ is the number density in the bulk and $\zeta$ is the drag coefficient.
For ABPs the orientation $\bq$ is governed by unbiased rotational Brownian diffusion  $D_R=1/\tau_R$. The swim diffusivity $D^{swim}=U_0^2\tau_R/6$ is isotropic. In analogy to the osmotic pressure of passive Brownian particles $\bsigma^{osmo}=-n\zeta D_T\bI=-nk_BT\bI$, where $D_T$ is the thermal translational Brownian diffusivity, we define $k_sT_s=\zeta U_0^2\tau_R/6$ in 3D and $k_sT_s'=\zeta U_0^2\tau_R/2$ in 2D \cite{Pressure2014,ForceBoundary2015}.

However, if one biases the orientation with an external field along some direction $\hat{\bH}$, then the swim diffusivity $\bD^{swim}$ is in general anisotropic. It is natural to keep the definition of swim stress as a confinement stress, $\bsigma^{swim}=-n\zeta \bD^{swim}$, but whether this definition is self-consistent in the mechanical sense is not known. In this work we address this question: can the swim stress be a true tensorial stress?

\begin{figure}[t]
	\centering
	\includegraphics[width=0.5\linewidth]{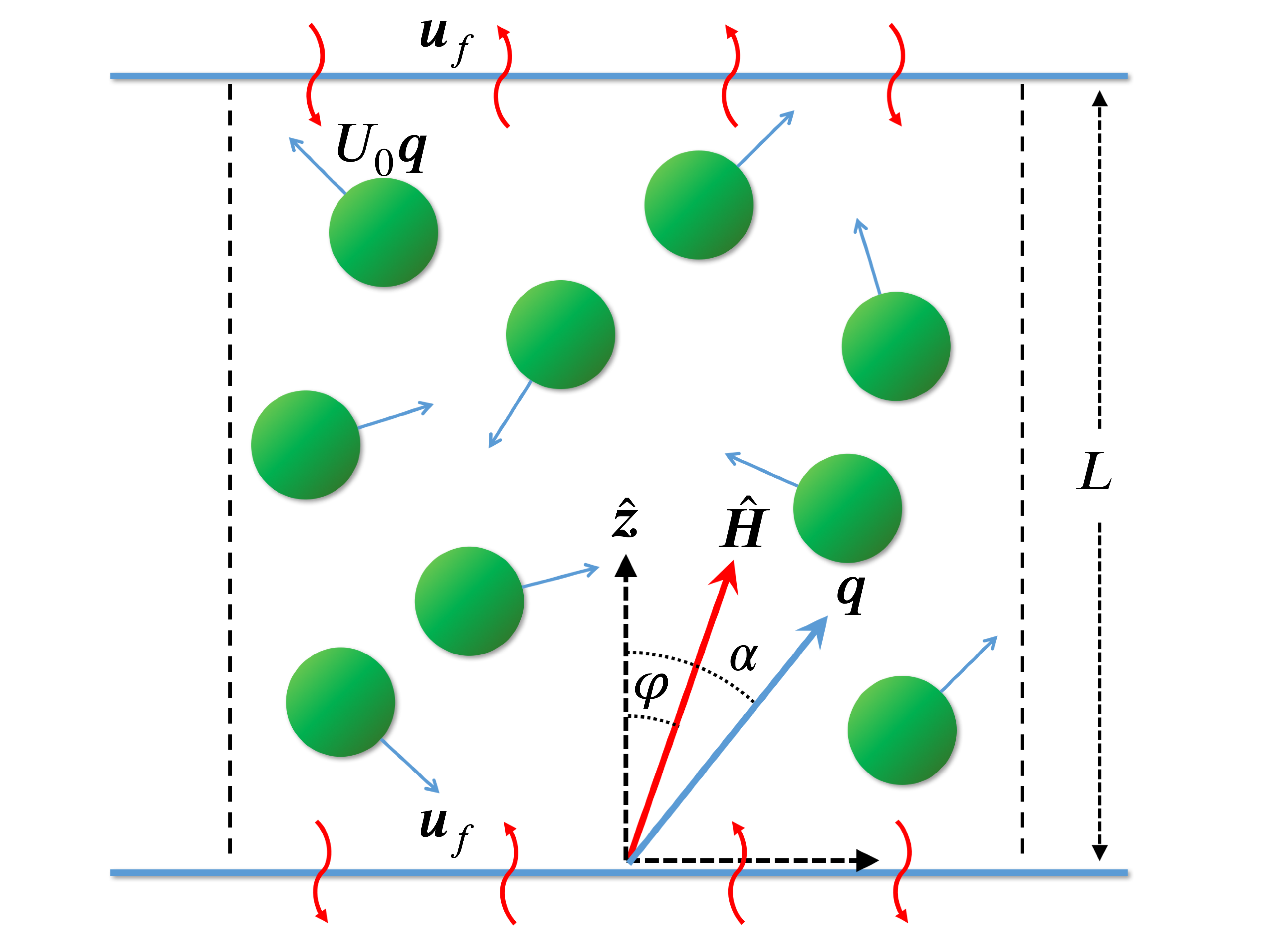}
	\caption{Sketch of the 2D geometry. The external field direction is $\hat{\bH}$ and $U_0\bq$ is the swim velocity of each ABP. The normal vector of the bottom wall $\bn=\hat{\bz}$, $\bq\cdot\hat{\bz}=\cos\alpha$, and $\hat{\bH}\cdot\hat{\bz}=\cos\varphi$.}
	\label{fig:geometry}
\end{figure}

Without loss of generality, we consider 2D ABPs between two parallel walls separated by $L$ as shown in Fig.~\ref{fig:geometry} under a bistable orientational potential energy function $V(\bq) = - \epsilon ( \bq\cdot\hat{\bH}  )^2$, 
where $\epsilon$ is an energy scale. Energy is minimized for  $\bq=\pm\hat{\bH}$; such a potential is  seen for magnetic nanoparticles \cite{coffey_thermal_2012}. We define $\chi_R=\epsilon/k_BT$ as the dimensionless strength of the field. 
The nematic field direction is applied at an angle $\varphi$ relative to the wall normal vector $\bn$: $\cos\varphi=\bn\cdot\hat{\bH}$.
Fluid is assumed to flow freely across the wall---it is an osmotic barrier---so that only the particle pressure is measured, and the wall-particle interaction is taken to be excluded-volume only.
The configuration is similar to the sedimentation problem \cite{SwimForce2015}, except that in this work we consider the dilute limit so swimmer-swimmer interactions are  ignored.
We also impose $L/\ell\to\infty$ to eliminate any confinement effects \cite{Ezhilan2015,ForceBoundary2015}, where the run-length $\ell=U_0\tau_R$.  

Note that this potential creates no polar order, $\langle \bq \rangle = 0$, only nematic order.
Further, since the orienting field is applied homogeneously and the wall-particle interactions are excluded volume only, issues associated with the force on a boundary differing from the bulk swim pressure \cite{Solon2015a} do not apply.  Thus, for these conditions, the tensorial continuum perspective predicts a normal  pressure on the wall  from the anisotropic swim stress: $\Pi^{W,swim} = -\left(\bsigma^{swim}\cdot\bn\right)\cdot\bn$.  

Independent of the  continuum perspective, the swim pressure on a wall has also been explained  microscopically \cite{ForceBoundary2015,CurvedBoundary2018} where ABPs accumulate in a wall boundary layer with thickness of order $\delta=\sqrt{D_T\tau_R}$. This colloidal perspective predicts $\Pi^{W,swim}=\Pi^{W,tot}-\Pi^{W,osmo}=(n^W-n)\zeta D_T$, where $n^W$ and $n$ are the number density of ABPs at the wall and in the bulk, respectively.

In this work we first follow the tensorial continuum perspective to calculate $\Pi^{W,swim}$ analytically.  We then use the colloidal perspective to calculate $\Pi^{W,swim}$ by solving the Smoluchowski equation for the distribution $P(\bx,\bq)$ at steady state, utilizing $n^W=\int P(\bx^W,\bq)d\bq$.
We show that the two perspectives agree with each other for arbitrary field direction $\hat{\bH}$, and also agree with the  force/area determined directly from Brownian dynamics (BD) simulations.
We further show that the normal stress difference generates a net flow of ABPs along the wall.

\textbf{The tensorial continuum (macroscopic) perspective.}
The swim stress $\bsigma^{swim}$ is an intrinsic property of ABPs in the bulk, regardless of the presence of a boundary. Therefore we consider only the relation between $\bq$ and $\hat{\bH}$ and define $\cos\theta=\bq\cdot\hat{\bH}, \theta\in[-\pi,\pi)$ for convenience.
At steady state, the equilibrium distribution $P_0^{\infty}$ of $\theta$ is:
\begin{align}
P_0^{\infty}(\theta) = e^{\frac{1}{2} \chi_R \cos (2 \theta)}/\left[2 \pi  I_0(\chi_R/2)\right].
\end{align}
The fluctuation of the orientation $\bq$ can be described by the $\bB$-field, which is analytically solvable with the Generalized Taylor Dispersion Theory \cite{FrankelGTDT1989}. In the weak and strong field limits ($\chi_R\to0,\infty$), the  behaviors can be calculated with a regular expansion or Kramer's hopping theory, respectively. The detailed solution and asymptotics are in the appendices:
\begin{subequations}
	\begin{align}
	B_\parallel(\theta) &= - \int_0^{\theta}\frac{\sqrt{\pi } e^{\chi_R \sin ^2\kappa} \ERF\left(\sqrt{\chi_R} \sin \kappa\right)}{2 \sqrt{\chi_R}}   d\kappa, \\
	B_\perp(\theta) &= \int_0^{\theta} \frac{F_D\left(\sqrt{{\chi_R}} \cos \kappa\right)}{\sqrt{\chi_R}} d\kappa,
	\end{align}
\end{subequations}
where $F_D(z)$ is the Dawson-$F$ integral function: $F_D(z)=e^{-z^2} \int_0^z e^{y^2} dy$.

\begin{figure}
	\centering
	\includegraphics[width=0.5\linewidth]{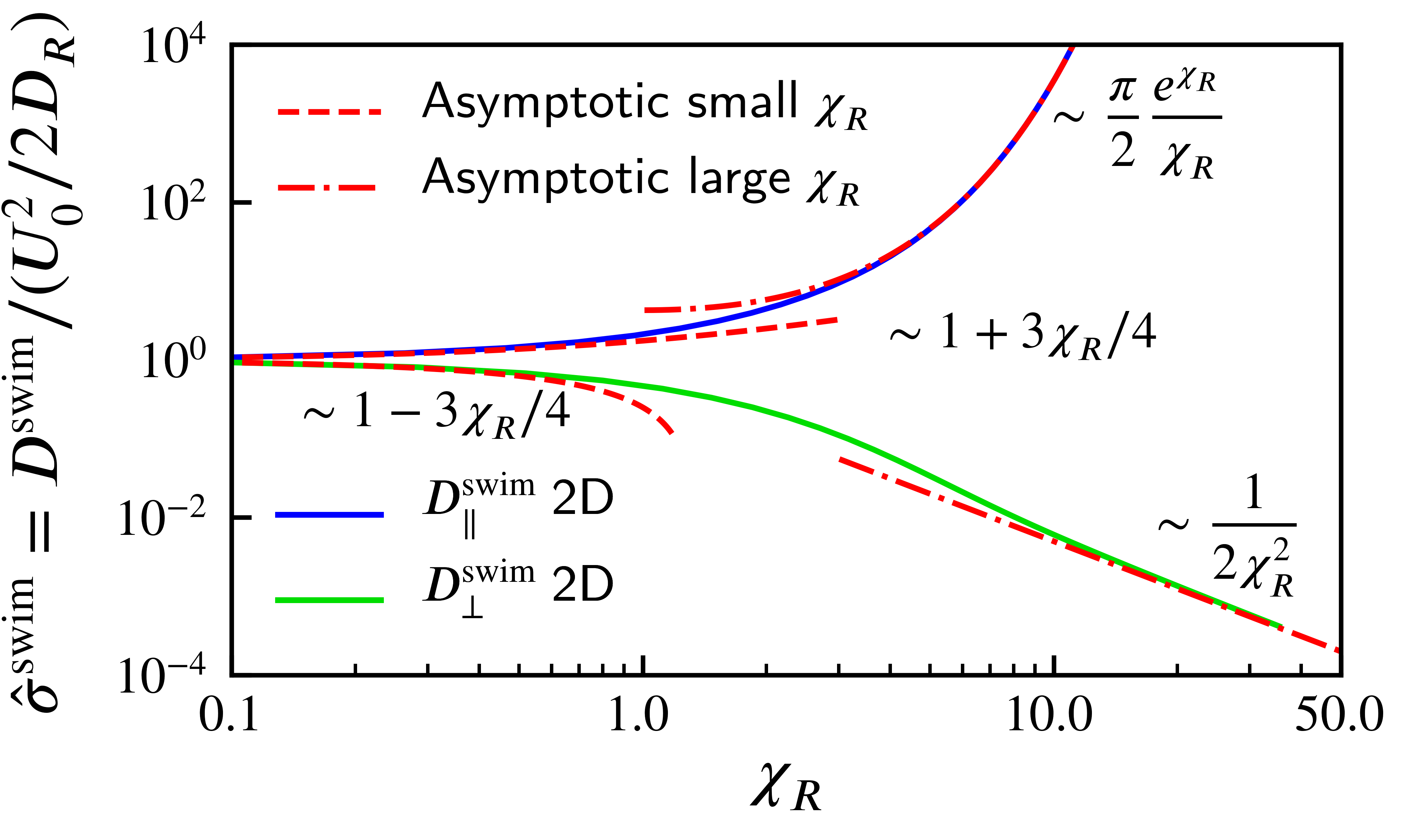}
	\caption{The swim diffusivity $\bD^{swim}$ in the directions parallel and perpendicular to the external field $\hat{\bH}$ in 2D. The solid lines are the analytical solutions~(\ref{eq:Dnemaswim2D}).}
	\label{fig:Diff2D}
\end{figure}

The swim diffusivity is generated by the orientational fluctuation $\bB$, propagated from the $\bq$ space to the $\bx$ space by the swim velocity $U_0\bq$:
\begin{subequations}\label{eq:Dnemaswim2D}
	\begin{align}
\hat{\sigma}_\parallel^{swim} = \frac{D_\parallel^{swim}}{U_0^2/2D_R} &= 2 \int_{-\pi}^{\pi} B_\parallel\left(\theta\right)P_0^{\infty}(\theta)\cos\theta d\theta, \\
\hat{\sigma}_\perp^{swim} = \frac{D_\perp^{swim}}{U_0^2/2D_R} &= 2\int_{-\pi}^{\pi} B_\perp\left(\theta\right)P_0^{\infty}(\theta)\sin\theta d\theta.
	\end{align}
\end{subequations}
The double integrals are numerically integrated and shown in Fig.~\ref{fig:Diff2D}.
Here $\hat{\sigma}_\parallel^{swim}$ and $\hat{\sigma}_\perp^{swim}$ are dimensionless functions representing the effects of $\chi_R$:
\begin{align}
	\frac{\bsigma^{swim}}{-nk_sT_s'} = \hat{\sigma}_\parallel^{swim} \hat{\bH}\hat{\bH} + \hat{\sigma}_\perp^{swim}\hat{\bH}_\perp \hat{\bH}_\perp\ ,
\end{align}
where $\hat{\bH}_\perp \hat{\bH}_\perp = \bI - \hat{\bH}\hat{\bH}$.

 In the limit $\chi_R\to\infty$ the diffusivities and stresses are very anisotropic: 
\begin{align}
 \hsigma_{\parallel}^{swim} \sim  \frac{\pi}{2}\frac{e^{\chi_R}}{\chi_R}, \quad \hsigma_{\perp}^{swim} \sim\frac{1}{2\chi_R^2};
\end{align}
the anisotropy, $\hsigma_{\parallel}^{swim}/\hsigma_{\perp}^{swim}\sim\pi\chi_R e^{\chi_R}$, grows exponentially with the field strength $\chi_R = \epsilon/k_BT$.  The exponential growth reflects the Kramer's hopping process:  at high $\chi_R$ a particle is trapped in either the  $\pm \hat{\bH}$ direction and requires a thermal fluctuation in orientation in order to overcome the barrier and flip to the other direction.

From the tensorial continuum perspective, we can {\em analytically} calculate the pressure on the wall for any $\varphi$:
\begin{align}\label{eq:pwalltensorial}
\frac{\Pi^{W,swim}}{n k_sT_s'} & = \hsigma_\parallel^{swim}(\hat{\bH}\cdot\bn)^2 + \hsigma_{\perp}^{swim}(\hat{\bH}_\perp\cdot\bn)^2.
\end{align}

\textbf{The colloidal (microscopic) perspective.}
Owing to symmetry, we only need to solve the Smoluchowski equation in the domain $z\in[0,L],\alpha\in[-\pi,\pi]$, with the boundary conditions being non-penetrating at $z=0,L$ and periodic in $\alpha$. The angle $\varphi$ is a parameter.  
All lengths are nondimensionalized with $\ell=U_0\tau_R$, and time is scaled with $\tau_R$; thus,
\begin{align}\label{eq:smolucolloid}
\dpone{P}{t}& + \dpone{}{z}j_T + \dpone{}{\alpha} j_R = 0,
\end{align}
where $j_T = \cos\alpha P - \left(\delta/\ell \right)^2 {\partial P}/{\partial z}$ and $j_R = \chi_R \sin 2(\alpha-\varphi) P - {\partial P}/{\partial \alpha}$.
These equations can be easily solved with a Finite Element PDE solver with non-penetrating boundary conditions on the top and bottom walls as illustrated in Fig.~\ref{fig:geometry}. We used the software package FreeFEM++ with automatic mesh refinement. After the steady state is reached, the swim pressure on the wall can be calculated as: $\Pi^{W,swim} = (n^W-n) \zeta D_T$, $n^W=\int P(z=0,\alpha) d \alpha$.

In addition to the Smoluchowski colloidal perspective of the swim pressure, we also perform BD simulations to verify both the colloid  and  continuum tensorial perspectives.
In the BD simulations, the pressure is determined directly as a summation of all the forces exerted by each particle-wall collision.
In both cases we set $n$ as the number density in the center of the channel; since the channel is wide enough to eliminate confinement effects $n$ is the uniform {\em bulk} number density used in the continuum derivation of the swim diffusivity and pressure.

\begin{figure}[t]
	\centering
	\includegraphics[width=0.5\linewidth]{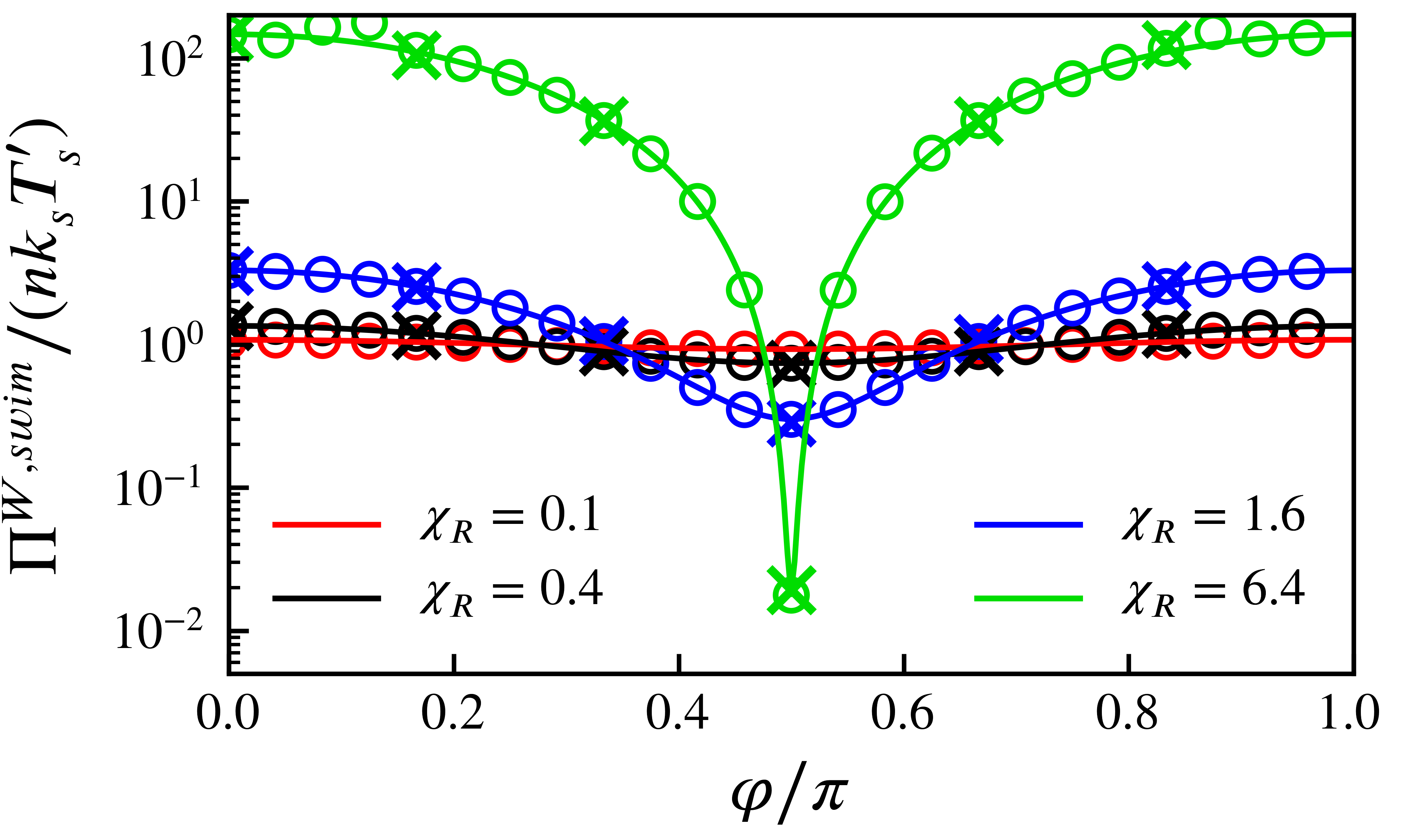}
	\caption{The comparison between the colloidal perspective, the continuum mechanics tensorial perspective, and Brownian dynamics simulations. The solid lines are calculated analytically from~(\ref{eq:pwalltensorial}), the open symbols are calculated from~(\ref{eq:smolucolloid}) via a FEM solver with $(\delta/\ell)^2=0.2$ and $L=20\ell$. The cross symbols are measured from particle-wall collisions in Brownian dynamics simulations with $\ell=4a$ and $\delta=0$ in a box with $L=128a=32\ell$ and periodic in the horizontal direction.}
	\label{fig:Pnorm}
\end{figure}

The comparison of the two different perspectives, together with results of Brownian dynamics simulations, is shown in Fig.~\ref{fig:Pnorm}.
All three methods agree with each other.   

The pressure calculated by~(\ref{eq:pwalltensorial}) is analytic and is valid for arbitrary ratio of swimming to diffusion, $\ell/\delta \in (0,\infty)$, and is also independent of the channel width $L$ as long as no confinement effects are important, i.e. $\ell/L \ll 1$.  The pressure from the colloid perspective is calculated for $(\delta/\ell)^2=0.2$, and $L=20\ell$ to guarantee that there are no confinement effects \cite{ForceBoundary2015}. The Brownian dynamics simulations are conducted with $D_T=0$.

The comparison clearly shows that the mechanical swim pressure on a wall satisfies the requirement of continuum mechanics, even when it is strongly anisotropic as shown for the case $\chi_R=6.4$.

\textbf{The tangential component $(\bsigma^{swim}\cdot\bn)\cdot\bt$.}
In continuum mechanics, $\bsigma^{swim}\cdot\bn$ is the traction on a plane with normal $\bn$, and the tangential component $(\bsigma^{swim}\cdot\bn)\cdot\bt$ in the tangential direction $\bt$ is the shearing force applied on that plane, i.e., the friction between the two continuous media.
For an anisotropic $\bsigma^{swim}$, the tangential component is not necessarily zero.

However, there cannot be any shear force (friction) in the ABP model because the wall-swimmer interaction is excluded volume only; that is, a force is transmitted only in the normal direction to prevent the swimmer from crossing the wall.  (In the ABP model hydrodynamics are neglected and thus there is no shear stress in the fluid.) When a swimmer swims towards a wall, it is trapped on the wall until the orientation $\bq$ relaxes to a different direction so that it can leave the wall. In the absence of friction, the tangent component of ABP's motion $U_0\bq\cdot\bt$ is not transmitted to the wall; the swimmer `slides' along the wall.
Therefore, the tangential component of swim stress results in a net boundary {\em flow} of ABPs along the wall.
The direction of the net flow is towards the left on the bottom wall and towards the right on the top wall for the $\hat{\bH}$ shown in Fig.~\ref{fig:geometry}.
The flow on the bottom and  top walls are the same magnitude but in opposite directions, and they cancel each other so there is no net overall motion in the domain and no net polar order.

For the 2D geometry shown in Fig.~\ref{fig:geometry}, the continuum tensorial stress perspective predicts the flow:
\begin{align}\label{eq:jsheartensorial}
\frac{\zeta \int\bj_T^{W,swim} dz \cdot\bt }{n k_sT_s'} = \left(\hsigma_\parallel^{swim} -  \hsigma_{\perp}^{swim}\right)\cos\varphi\sin\varphi.
\end{align}
It is clear that if $\chi_R=0$, $\hsigma_\perp^{swim}=\hsigma_\parallel^{swim}$ and the flow varnishes; only `normal stress differences' drive a flow. Here, $\int\bj_T^{W,swim} dz$ has the dimension of the total flow rate along the boundary per unit boundary length (area if in the 3D case), while $\zeta \int\bj_T^{W,swim} dz \cdot\bt $ has the dimension of pressure.

Note that the stress difference driving the boundary flow is actually the total stress difference, $\sigma^{tot}_\parallel - \sigma^{tot}_\perp$, where $\bsigma^{tot}=\bsigma^{swim}+\bsigma^{osmo}$, since the osmotic pressure is isotropic and cannot generate a normal stress difference.

From the microscopic colloid perspective, the swimmers form a kinetic boundary layer \cite{ForceBoundary2015} on the wall. More specifically, there is net polar order $\bmm=\int P(\bx,\bq)\bq d\bq\neq 0$ in the boundary layer close to the wall, even though the nematic orientation field has no polar order in the bulk. By solving the Smoluchowski equation~(\ref{eq:smolucolloid}), the flow is obtainable by integrating $m_t$, the component of $\bmm$ parallel to the wall:
$\int\bj_T^{W,swim} dz \cdot\bt = U_0 \int \bmm\cdot\bt dz$. More details about this boundary layer can be found in the appendices.
\begin{figure}
	\centering
	\includegraphics[width=0.5\linewidth]{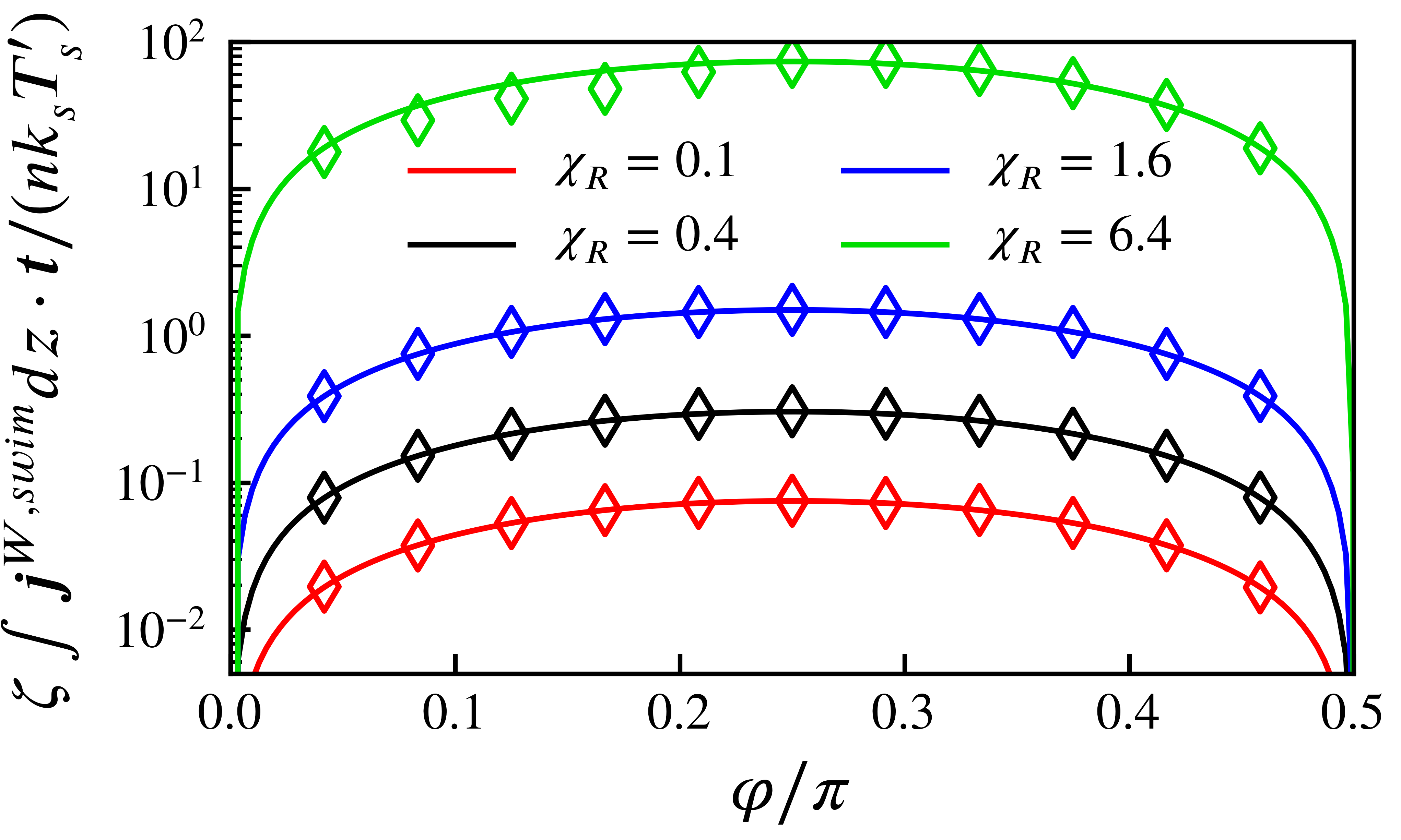}
	\caption{Comparison between the colloidal perspective and the continuum mechanics tensorial perspective for the shear component and the corresponding flow along the wall. The solid lines are calculated analytically from~(\ref{eq:jsheartensorial}); the open symbols are data calculated with the colloidal perspective $U_0\int\bmm\cdot\bt dz$ from the same FEM solution as in Fig.~\ref{fig:Pnorm}.}
	\label{fig:Pshear}
\end{figure}
The comparison between the colloidal perspective and the continuum mechanics tensorial perspective  (\ref{eq:jsheartensorial}) for the tangential component, and the corresponding flow along the wall is shown in Fig.~\ref{fig:Pshear}.  The agreement is excellent.

\textbf{Conclusions \& Discussion.}
In this work we presented an example designed to extend the notion of the swim pressure to a true tensorial swim stress for the case of swimmers in a nematic orientation field.
Swimmers under a nematic orientational potential show dramatically enhanced diffusion parallel to the field direction $\hat{\bH}$, and significantly reduced diffusivity in the $\hat{\bH}_\perp$ direction. 
This is in contrast to the polarization case \cite{Polar2014,SwimForce2015}, where all swimmers are directed towards the same direction and the diffusivity in both the $\hat{\bH}$ and $\hat{\bH}_\perp$ directions decay algebraically with increasing $\chi_R$. 

The anisotropic swim diffusivity gives an anisotropic swim stress from the general relation between diffusion and stress: $\bsigma^{swim}=-n\zeta\bD^{swim}$. 
Using a parallel-wall geometry, we showed that an anisotropic swim stress is a true stress in the continuum mechanical sense---the pressure on a boundary is  $\Pi^{W,swim}=-(\bsigma^{swim}\cdot\bn)\cdot\bn$. This applies for an isotropic state, a state with polar order or one with nematic order. 

In the absence of hydrodynamics, the tangential component of the anisotropic swim stress does not generate a shear stress (friction)  but rather a net flow of ABPs along the wall. This is because the interaction between the ABP and the wall is assumed to be frictionless. 
From the continuum perspective the flow along the boundary is driven by normal stress differences.  This is a generic feature of active matter systems.  Due to confinement \cite{ForceBoundary2015} or an orienting field, the stress in active matter is anisotropic.  If the boundary orientation does not coincide with the principle axes of the swim stress tensor, net boundary flow will result.  In the presence of hydrodynamics, this flow of swimmers along the wall would  drag  fluid with it and result in a shear stress. From the continuum perspective, the swim stress $\bsigma^{swim}$ contributes no shear component to the total stress of suspension $\bsigma^{sus}$, and so the flow must be balanced by the viscous shear stress $2 \eta \be$, where $\be=\tfrac{1}{2}\left(\nabla \bu +\nabla\bu^T\right)$, with $\bu$ the suspension average velocity and $\eta$ is the fluid viscosity. The vorticity generated by the fluid shear stress will affect the orientation distribution near the wall and thus the anisotropic swim stress.  This effect is left for a future study.

It is unclear at this stage whether the swim stress can  be treated generally as a true tensorial stress for arbitrary externally imposed orientational motion beyond the nematic ordering case discussed in this work.
Rigorous mathematical proof requires solution of the kinetic boundary layer with arbitrary orientation order and is usually difficult. The orientation moment expansion method \cite{Saintillan2015,ForceBoundary2015, CurvedBoundary2018} may be a possible route towards a general proof, but it is subject to proper orientation closure relations.
We leave this for a future study. 

The continuum tensorial perspective of swimmers has more profound use than simply to estimate the pressure on a flat wall for swimmers without net motion. Although there have been some debates about whether the anisotropic swim stress can be an equation of state \cite{Solon2015a,Hagen2015,Ginot2015,Junot_Briand_Ledesma-Alonso_Dauchot_2017}, in this work we showed that from a purely mechanical perspective the anisotropic swim stress can be self-consistent and useful in predicting the surface forces. In a general mechanical transport problems such as sedimentation or active micro-rheology, the large scale motion and deformation of swimmers can be simply solved with the continuum mechanics flux $\bj_{cm}$ driven by the tensorial stress, body force, and swim force \cite{SwimForce2015}:
\begin{align}
\bj_{cm} = \frac{1}{\zeta} \left(\nabla\cdot\bsigma^{tot} + n\bF^g + n\AVE{\bF^{swim}} \right).
\end{align}   
The boundary conditions for this large-scale transport equation must be properly constructed from the detailed near-wall dynamics on the small scale.
This is similar to rarefied gas dynamics, where the non-continuum effects must be resolved on the scale of a few mean free paths at the boundary, and then a proper boundary condition for Navier-Stokes equation in the outer region can be constructed from the `inner' solution.
A similar outer-inner matching scheme also applies for ABPs, as discussed in our previous work on the curved kinetic boundary layer of active matter \cite{CurvedBoundary2018}.

\begin{acknowledgments}
This work is supported by NSF CBET-1437570.
\end{acknowledgments}

\appendix

\section{Generalized Taylor Dispersion Theory}
\label{sec:app_aniso2d}
In this section we follow the Generalized Taylor Dispersion Theory (GTDT) by  \citet{FrankelGTDT1989} to derive the anisotropic swim diffusivity $\bD^{swim}$ and the ideal gas swim stress $\bsigma^{swim} = - n\zeta \bD^{swim}$. Similar methods have also been used by  \citet{Zia2010} and by  \citet{Polar2014}.
In the $\bB$-field theory by  \citet{FrankelGTDT1989}, $\bq$ is a local degree of freedom. For the swimmers considered here, $\bq$ is the orientation vector of each swimmer.
The steady state distribution, $P_0^\infty(\bq)$, is analytically solvable from the balance of rotational flux $\bj_R$:
\begin{equation}
\bj_R = \bomega(\bq;\hat{\bH})P - \bD_R\cdot \nabla_R P,\quad\nabla_R\cdot\bj_R=0,
\end{equation}
where $\hat{\bH}$ is the unit vector in the direction of the orienting field, $\bomega(\bq;\hat{\bH})$ is the angular velocity. $\bD_R$ is the \emph{intrinsic} rotational diffusivity, which could be an anisotropic tensor.

The orientation-average velocity is defined as:
\begin{equation}
\langle\bU\rangle=\int_{\bq}P_0^\infty(\bq) \bU(\bq) d\bq.
\end{equation} 

By decomposing  $\Delta\bU(\bq) = \bU(\bq) - \langle\bU\rangle $,
the effective diffusivity is given by
\begin{equation}
\bD^{swim} = \int_{\bq} P_0^\infty(\bq) \bB(\bq)\Delta\bU(\bq) d\bq,
\end{equation}
where the  $\bB$ field is the solution to
\begin{align}
	\nabla_{\bq} \cdot \left[ \bomega P_0^\infty\bB - \bD_R\cdot\nabla_{\bq} (P_0^\infty\bB) \right] &= \Delta\bU P_0^\infty,\\
	\int_{\bq} P_0^\infty \bB d\bq &=0,
\end{align}
with appropriate BCs in $\bq$ space. Here $\bomega$ and $\bD_R$ are angular velocity and (intrinsic) rotational diffusivity in $\bq$ space, respectively. Physically, $\bB(\bq)$ represents the fluctuation of $\bq$ as a function of $\bq$. This fluctuation in the orientational space propagates to the translational motion physical space through the disturbance velocity $\Delta\bU(\bq)$.

For an orientational potential energy $V(\bq)$, the torque and angular velocity are:
\begin{align}
	\bL =& - \nabla_R V,\quad\quad \bomega = \frac{1}{\zeta_R} \bL,
\end{align}
where we assumed the isotropic orientational drag $\zeta_R$. 
The angular velocity is interpreted as:
\begin{align}
	\dot{\bq} = - \bq \times \bomega \, .
\end{align}

In this work we considered a special case where the potential energy $V(\bq) = - \epsilon (\bq \cdot\hat{\bH})^2$ is given by the bistable form in the main text. The direction of $V(\bq)$ is denoted by $\hat{\bH}$.  The parameter $\chi_R = \epsilon/k_BT$ sets the nondimensional strength of the potential.

\begin{figure}[t]
	\centering
	\includegraphics[width=0.5\linewidth]{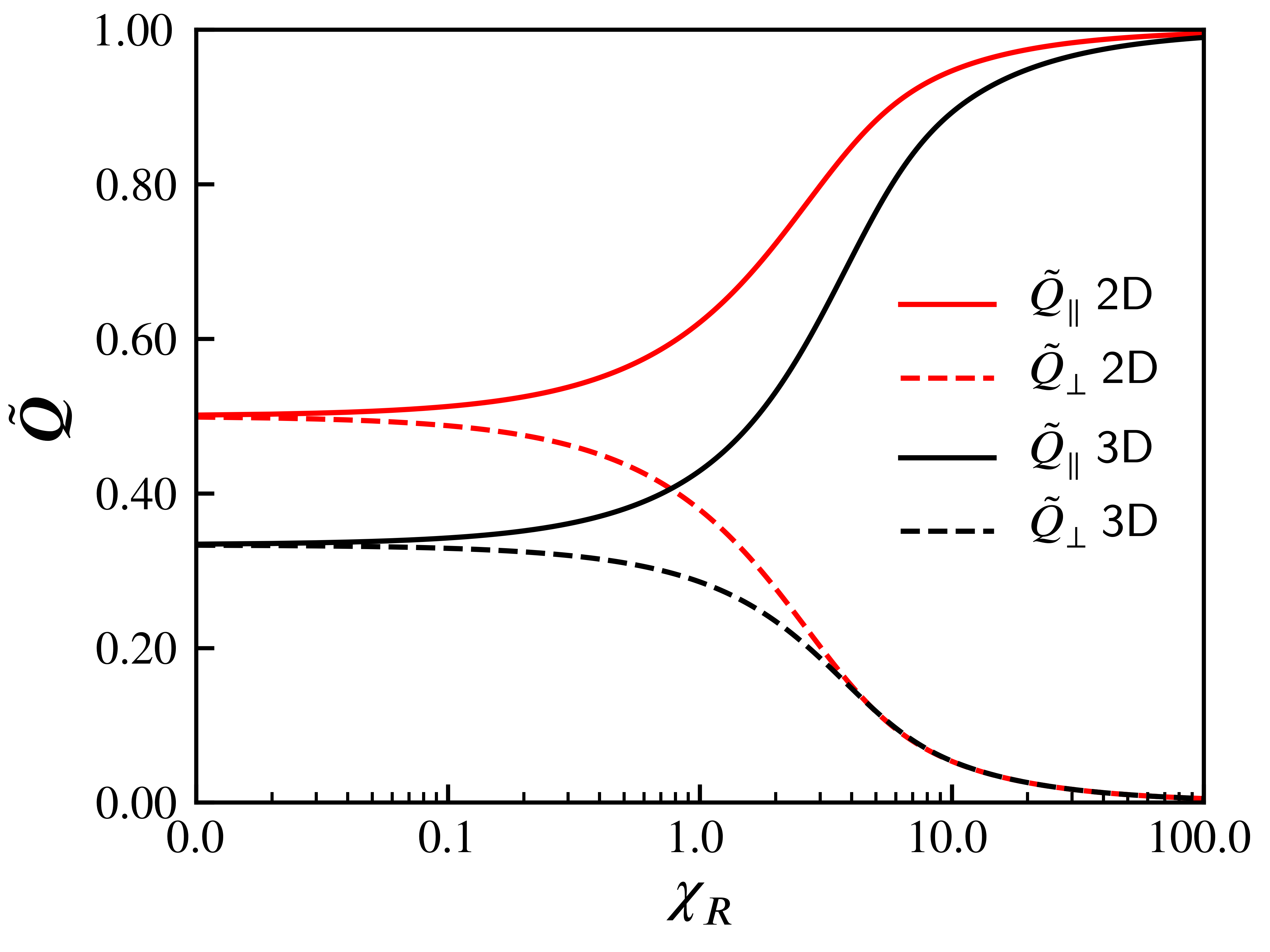}
	\caption{\label{fig:Q2D3D}The nematic order parameter $\tilde{\bQ}=\AVE{\bq\bq}$ as a function of field strength $\chi_R={\epsilon}/{k_BT}$.}
\end{figure}

\section{Case 1. Swimmers in a 2D layer: in-plane rotation.}
The rotational space for in-plane rotation is represented by a single angle $\theta\in[-\pi,\pi)$.
Let $\cos\theta=\bq\cdot\hat{\bH}$.
At steady state, the equilibrium orientation distribution is:
\begin{align}
	P_0^\infty(\theta) = \frac{1}{2 \pi  I_0(\chi_R/2)}e^{\frac{1}{2} \chi_R \cos (2 \theta)},
\end{align}
where $I_0$ is the Bessel function. $P_0^\infty(\theta)$ is normalized so that $\int_{-\pi}^\pi P_0^{\infty} d\theta = 1$.
The nematic order parameter $\tilde{\bQ}$ is:
\begin{subequations}
	\begin{align}
		\AVE{\bq_\parallel\bq_\parallel} &= \frac{1}{2} \left(\frac{I_1(\chi_R/2)}{I_0(\chi_R/2)}+1\right),\\
		\AVE{\bq_\perp\bq_\perp} &= \frac{1}{2} \left(- \frac{I_1(\chi_R/2)}{I_0(\chi_R/2)}+1\right).
	\end{align}
\end{subequations}
Here we also have $\Tr\tilde{\bQ}=1$, as required by the definition of $\tilde{\bQ}$. The zero-traced nematic order parameter $\bQ$ is defined as $\bQ=\tilde{\bQ}-\tfrac{1}{2}\bI$ for the 2D case.
The order parameter $\tilde{\bQ}$ is shown in Fig.~\ref{fig:Q2D3D}.

GTDT gives the $\bB$ field in the two directions parallel and perpendicular to $\hat{\bH}$:
\begin{subequations}
	\begin{align}
		B_\parallel(\theta) &= - \int_0^{\theta}\frac{\sqrt{\pi } e^{\chi_R \sin ^2\kappa} \ERF\left(\sqrt{\chi_R} \sin \kappa\right)}{2 \sqrt{\chi_R}}   d\kappa, \\
		B_\perp(\theta) &= \int_0^{\theta} \frac{F_D\left(\sqrt{{\chi_R}} \cos \kappa\right)}{\sqrt{\chi_R}} d\kappa,
	\end{align}
\end{subequations}
where $F_D(z)$ is the Dawson-$F$ integral function:
\begin{equation}
F_D(z)=e^{-z^2} \int_0^z e^{y^2} dy.
\end{equation}
The swim diffusivity comes from the orientational fluctuation $\bB$:
\begin{subequations}\label{eq:Dnemaswim2Dapp}
	\begin{align}
		\hat{D}_{\parallel}^{swim} = \frac{D_\parallel^{swim}}{U_0^2/2} &= - 2 \int_{-\pi}^{\pi} \int_0^{\theta}\frac{\sqrt{\pi } e^{\chi_R \sin ^2\kappa} \ERF\left(\sqrt{\chi_R} \sin \kappa\right)}{2 \sqrt{\chi_R}} d\kappa P_0^{\infty}(\theta)\cos\theta d\theta, \\
		\hat{D}_{\perp}^{swim} = \frac{D_\perp^{swim}}{U_0^2/2} &= 2\int_{-\pi}^{\pi} \int_0^{\theta} \frac{F\left(\sqrt{{\chi_R}} \cos \kappa\right)}{\sqrt{\chi_R}} d\kappa P_0^{\infty}(\theta)\sin\theta d\theta,
	\end{align}
\end{subequations}
which are shown in the maintext.

The swim stress follows
\begin{align}
	\hsigma_{\parallel}^{swim} &= \frac{\sigma_{\parallel}}{-n\zeta U_0^2/2}=\hat{D}_{\parallel}^{swim},\\
	\hsigma_{\perp}^{swim} &= \frac{\sigma_{\perp}}{-n\zeta U_0^2/2}=\hat{D}_{\perp}^{swim};
\end{align}
for 2D in-plane rotations the isotropic swim pressure is $n\zeta U_0^2/2$, instead of $n\zeta U_0^2/6$.

\subsection{The weak field limit $\chi_R\to 0$. }
By direct expansion of~(\ref{eq:Dnemaswim2Dapp}):
\begin{subequations}\label{eq:Dnemaswim2dweak}
	\begin{align}
		\hsigma_{\parallel}^{swim} &\approx 1+\frac{3 \chi_R}{4}+O(\chi_R^2),\\
		\hsigma_{\perp}^{swim} &\approx 1-\frac{3 \chi_R}{4}+O(\chi_R^2).
	\end{align}
\end{subequations}

\subsection{The strong field limit $\chi_R\to \infty$. }
In this case Kramers' escape rate theory can be directly used since the orientation is a 1D space for $\theta$.
For the potential $V(\theta)$, the escape rate out of its minimum is
\begin{align}
	r_K &= \frac{1}{2\pi} \sqrt{{V''(\theta_{min})}\ABS{V''(\theta_{max})}} e^{-\frac{V(\theta_{max})-V(\theta_{min})}{\zeta D}}\nonumber\\
	&= \frac{\chi_R {D_R}}{\pi } e^{-\chi_R} ,
\end{align}
where $V(\theta_{min})$ and $V(\theta_{max})$ are minimum and maximum of the potential $V$, respectively. 
The parallel swim diffusivity $D^{swim}_{\parallel}$ is the result of the 1D random walk in the direction of $\hat{\bH}$, and
\begin{align}
	\hsigma_{\parallel}^{swim}= \frac{D_\parallel^{swim}}{U_0^2\tau_R/2} \to\frac{\pi}{2} \frac{e^{\chi_R}}{\chi_R}.
\end{align}

The limiting transverse diffusivity $D^{swim}_{\perp}$ results from a `boundary layer' around the equilibrium position $\bq\cdot\hat{\bH}=0$, since at the strong field limit $\theta\approx 0$ (or $\pi$) is almost always true.
For 2D rotation, it can be directly calculated from the integral with the `boundary layer' approximation: $\theta\approx\sin\theta, \ \cos\theta\approx 1-\theta^2/2$.
The integrals in~(\ref{eq:Dnemaswim2Dapp}) are explicitly integrable with these approximations, and:
\begin{align}
	\frac{D_\perp^{swim}}{U_0^2\tau_R/2} &\approx 8 \int_0^{\frac{\pi }{2}} \frac{ e^{\frac{1}{2} \chi_R \cos (2 \theta)} \sin^2\theta }{4 \pi  \chi_R I_0\left(\frac{\chi_R}{2}\right)} \, d\theta = \frac{1-\frac{I_1\left(\frac{\chi_R}{2}\right)}{I_0\left(\frac{\chi_R}{2}\right)}}{2 \chi_R}
	\to \frac{1}{2 \chi_R^2 },
\end{align}
Therefore:
\begin{align}
	\hsigma_{\perp}^{swim}&=  \frac{D_\perp^{swim}}{U_0^2\tau_R/2} \to \frac{1}{2\chi_R^2}.
\end{align}
The asymptotics in the strong and weak limits are also shown in the main text.

\section{Case 2. Swimmers in 3D space. }
In this case the rotation is mathematically challenging to describe. In this work we follow the convention of  \citet{Brenner1972} by defining a nabla operator in orientation space $\nabla_R$. The evolution of a spherical ABP with orientation $\bq$ by torque and Brownian motion can be described in a spherical coordinate system  $(0<\theta<\pi,0<\phi<2\pi)$:
\begin{align}
	\bq = \sin\theta\cos\phi \be_x + \sin\theta\sin\phi \be_y + \cos\theta\be_z
\end{align}

The rotational gradient operator $\nabla_R = \bq\times\dpone{ }{\bq}$. Here we have:
\begin{align}
	\dpone{f(\theta,\phi)}{\bq} =& \be_{\theta} \dpone{f}{\theta} + \frac{1}{\sin\theta} \be_{\phi} \dpone{f}{\phi} \\
	\nabla_R = \bq\times\dpone{f}{\bq} =& \be_{\phi}\dpone{f}{\theta} - \frac{1}{\sin\theta} \be_{\theta}\dpone{f}{\phi} 
\end{align}
Also, the operators are usually used with its derivatives:
\begin{align}
	\dpone{ }{\bq} \bq =& \bI - \bq\bq \\
	\bq\cdot\dpone{ }{\bq} =& \dpone{ }{\bq}\cdot\bq= 0, \dpone{ }{\bq}\times\bq=0 \\
	\left(\bq\times\dpone{ }{\bq}\right) \times\bq=&-2\bq\\
	\bq\times\left(\bq\times\dpone{ }{\bq}\right) =&-\dpone{ }{\bq}\\
	\nabla_R\cdot\nabla_R =&\frac{1}{\sin\theta}\left(\dpone{ }{\theta}\sin\theta\dpone{ }{\theta}
	\right) + \frac{1}{\sin^2\theta}\dptwo{ }{\phi}
\end{align}

With these notations, the orientation is analyzed in the spherical coordinate system $\bq=(\theta,\phi)$, with $\theta\in[0,\pi],\phi\in[0,2\pi)$. 
The $\theta=0$ axis is chosen such that $\bq\cdot\hat{\bH}=\cos\theta$. 
The orientational distribution of $\bq$ obeys the Boltzmann distribution, regardless of the translational location $\bx$ of the swimmer:
\begin{align}
	P_0^\infty(d\Omega(\theta,\phi)) \propto \exp{\left(-V(\bq)/k_BT\right)} d\Omega ,
\end{align}
where $d\Omega$ is the solid angle. The equilibrium distribution is:
\begin{align}\label{eq:Peq3D}
	P_0^{\infty}\left( \theta,\phi \right) = \frac{\sqrt{\chi_R} e^{\chi_R}}{2 \pi ^{3/2} \ERFI\left(\sqrt{\chi_R}\right)} \exp{\left(-\chi_R \sin^2\theta\right)},
\end{align}
and $\phi$ does not appear due to the axisymmetry.
Here $\ERFI$ is the `imaginary error function', and $\chi_R = \dfrac{\epsilon}{k_BT}$ is the dimensionless field strength.
When $\chi_R=0$, the orientational potential $V$ vanishes and $P_0^\infty=1/4\pi$.

Due to the symmetry of the field $V(\bq)$, the polar order, $\langle \bq \rangle$,  is zero, and the effect of the field is quantified by the nematic order parameter $\tilde{\bQ}=\AVE{\bq\bq}$, as shown in Fig.~\ref{fig:Q2D3D}.
When $\chi_R=0$, $\tilde{Q}_{\perp}=\tilde{Q}_{\parallel}=1/3$. When $\chi_R\to\infty$, all particles with the field $\bq=\pm\hat{\bH}$, and therefore $\tilde{Q}_\parallel=1$ and $\tilde{Q}_{\perp}=0$:
\begin{subequations}
	\begin{align}
		\AVE{\bq_\parallel\bq_\parallel} &= \frac{\exp (\chi_R)}{\sqrt{\pi } \sqrt{\chi_R} \ERFI\left(\sqrt{\chi_R}\right)}-\frac{1}{2 \chi_R},\\
		\AVE{\bq_\perp\bq_\perp} &= \frac{1}{2} \left(1-\AVE{\bq_\parallel\bq_\parallel}\right).
	\end{align}
\end{subequations}
Here by definition $\Tr \tilde{\bQ}=1$. 
The zero-traced nematic order parameter $\bQ=\tilde{\bQ}-\tfrac{1}{3}\bI$, and $Q_\parallel=\AVE{\bq_\parallel\bq_\parallel}-1/3$, $Q_\perp=\AVE{\bq_\perp\bq_\perp}-1/3$.

\begin{figure}
	\centering
	\includegraphics[width=0.5\linewidth]{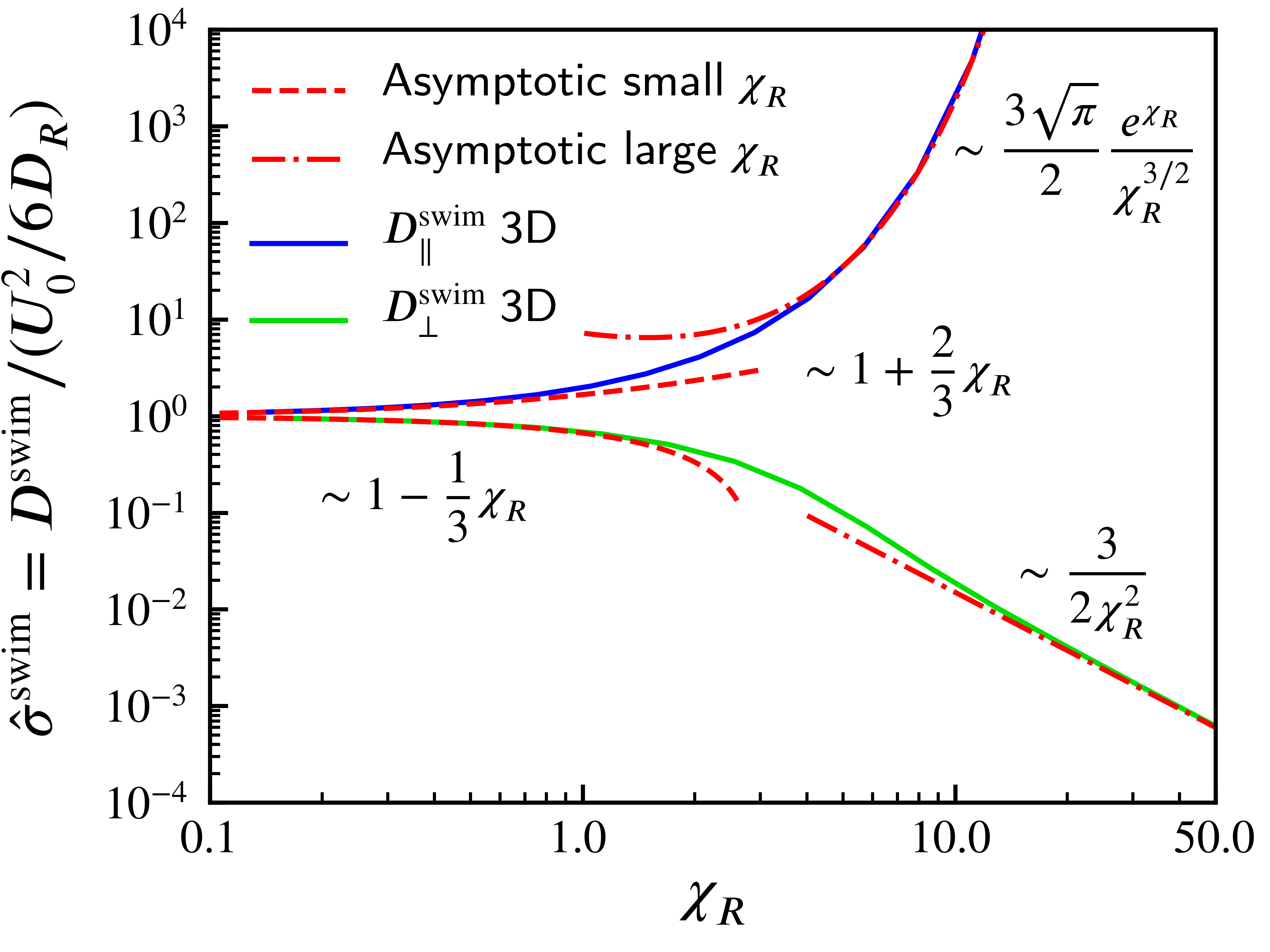}
	\caption{The swim diffusivity $\bD^{swim}$ in the directions parallel and perpendicular to the external field $\hat{\bH}$ in 3D space. The solid lines are the analytical solutions~(\ref{eq:Dnemaswim3d}).}
	\label{fig:Diff3D}
\end{figure}

The solution for $\bB(\bq)$ is:
\begin{subequations}
	\begin{align}
		B_{\parallel} (\theta) &= \int_0^{\cos\theta} \frac{1-e^{\chi_R-\chi_R k^2} }{2 \chi_R \left(k^2-1\right)} \, dk,\\
		B_{\perp}(\theta) &= \cos \phi \sin \theta g(\cos\theta),
	\end{align}
\end{subequations}
where the function $g(x)$ in $B_\perp$ is the solution of the ODE:
\begin{align}\label{eq:BperpG}
	&\left(x^2-1\right) g''(x)+2 x \left(\chi_R \left(x^2-1\right)+2\right) g'(x)\nonumber \\
	&+2 \left(\chi_R x^2+1\right) g(x)-1=0.
\end{align}
The function $g(x)=g(\cos\theta)$ satisfies (i) no singularities at $x=\cos\theta\to\pm1$, and (ii) is well-defined as $\chi_R\to 0$.
Thus,
\begin{subequations}
	\label{eq:Dnemaswim3d}
	\begin{align}
		\hat{D}_{\parallel}^{swim} = \frac{D_\parallel^{swim}}{ U_0^2\tau_R/6}&= 12\pi \int_0^{\pi} P_0^\infty \cos\theta\sin\theta \int_0^{\cos\theta} \frac{1-e^{\chi_R-\chi_R k^2} }{2 \chi_R \left(k^2-1\right)} \, dk \, d\theta, \\
		\hat{D}_{\perp}^{swim}=\frac{D_\perp^{swim}}{ U_0^2\tau_R/6}&= 6\int_0^{2\pi}\int_0^{\pi} P_0^\infty \sin^3\theta  g(\cos\theta) \cos^2\phi \, d\theta \, d\phi,
	\end{align}
\end{subequations}
which are shown in Fig.~\ref{fig:Diff3D}.

From the swim diffusivity the swim stress follows as $\bsigma^{swim}=-n\zeta \bD^{swim}$, and 
\begin{align}
	\hsigma_{\parallel}^{swim} &= \frac{\sigma_{\parallel}}{-n\zeta U_0^2/6}=\hat{D}_{\parallel}^{swim},\\
	\hsigma_{\perp}^{swim} &= \frac{\sigma_{\perp}}{-n\zeta U_0^2/6}=\hat{D}_{\perp}^{swim}.
\end{align}

\subsection{The weak-field limit $\chi_R\to 0$. }
For weak fields a regular expansion  of~(\ref{eq:Dnemaswim3d}) gives:
\begin{subequations}
	\begin{align}
		\hsigma_{\parallel}^{swim} &\approx 1+\frac{2\chi_R}{3}+O(\chi_R^2),\\
		\hsigma_{\perp}^{swim} &\approx 1-\frac{\chi_R}{3}+O(\chi_R^2).
	\end{align}
\end{subequations}
As was the case for polar order aligned with $\hat{\bH}$ induced by a potential with a single position of minimum energy  \cite{Polar2014}, the swim pressure is decreased in the $\hat{\bH}_\perp$ direction, because the energy barrier decreases the fluctuation of orientation $\bq$ in that direction.
The difference here, however, is that the stress in the $\hat{\bH}$ direction is enhanced by the field. This is due to the bistable structure of the orientation potential, and we shall see a more significant effect in the strong-field limit.

\subsection{The strong-field limit $\chi_R\to\infty$. }
In this case, the swimmers may all align with either $\hat{\bH}$ or $-\hat{\bH}$, and only occasionally `jump' between these two states.
This is analogous to the famous Kramers' escape  process   \cite{kramers_brownian_1940}, where a Brownian particle may jump out of a potential well slowly due to diffusion.
As $\bq$ is diffusive in rotation space, the jumping probability is modified from Kramers' original 1D estimation.
The average jump time $\tau_j$ between the two directions is estimated to be \cite{coffey_escape_2001}:
\begin{align}
	\tau_j= \frac{\sqrt{\pi } \exp (\chi_R)}{2 \chi_R^{3/2}} \tau_R.
\end{align}

Physically, the swimmer may move in a direction with $U_0$ for $\tau_j$ and then jump to the other direction and move again with $U_0$ for another $\tau_j$.
Therefore, at times long compared to $\tau_R$ {\em and} $\tau_j$, the diffusivity is simply a 1D random walk in the direction of $\hat{\bH}$: 
\begin{align}
	\hsigma_{\parallel}^{swim}=  \frac{D_\parallel^{swim}}{U_0^2\tau_R/6} \to \frac{3\sqrt{\pi } \exp (\chi_R)}{2 \chi_R^{3/2}}.
\end{align}
It is important to note that one must wait a time long compared to $\tau_j$ before the limiting behavior is obtained and this time grows exponentially with the field strength $\chi_R$.

In addition to  moving in the $\pm\hat{\bH}$ directions, the swimmers also move in the direction perpendicular to $\hat{\bH}$, due to small fluctuations around $\pm\hat{\bH}$ driven by $D_R$.
Following this route, the distribution of the fluctuation field $B_\perp$ can be approximated with a singular `boundary layer' around the parallel direction.
After the tedious mathematics is properly handled, the result is very simple:
\begin{align}
	\hsigma_{\perp}^{swim}= \frac{D_\perp^{swim}}{U_0^2\tau_R/6} \to \frac{3}{2\chi_R^2},
\end{align}
as $\chi_R\to\infty$. The asymptotic predictions are shown in Fig.~\ref{fig:Diff3D} and are in excellent agreement with the full solutions.

\section{The polar order in the boundary layer}
\begin{figure}[h]
	\centering
	\includegraphics[width=0.5\linewidth]{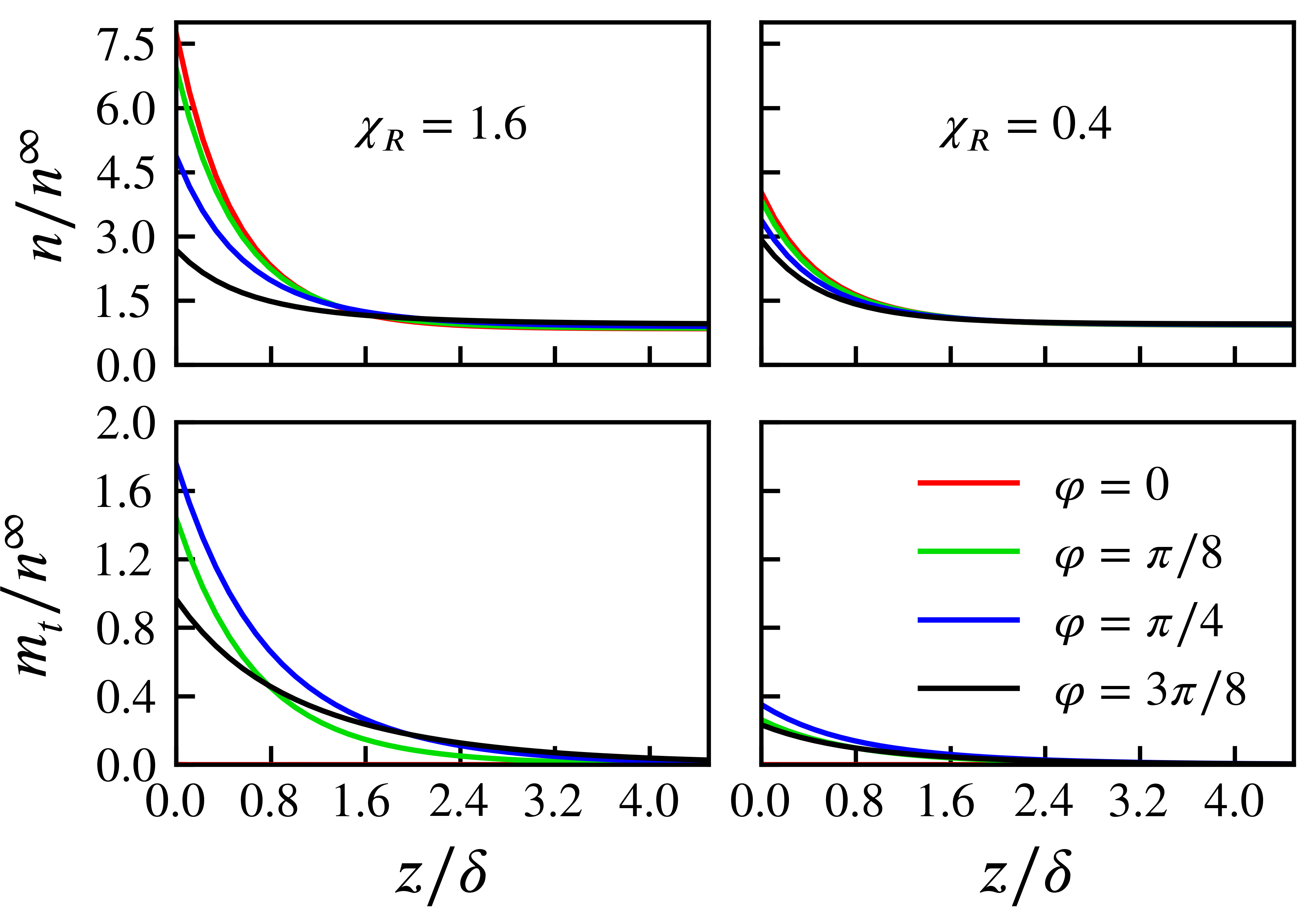}
	\caption{\label{fig:Fig_BLnm} The boundary-layer structure for the case of $\chi_R=1.6$ (left column) and $\chi_R=0.4$ (right column), taken from the same data as shown in Fig.~4 of the main text. Here $n^\infty$ is the number density in the bulk, corresponding to the $n$ in Fig.~3 and Fig.~4 of the main text. The boundary-layer thickness $z$ is scaled with the microscopic length $\delta = \sqrt{D_T\tau_R}$. The tangential component of polar order $m_t=\bmm\cdot\bt$.  For the $\hat{\bH}$ in Fig.~1 of the main text, $m_t$ is towards the left on the bottom wall. For $\varphi=0$, $m_t=0$ everywhere..}
\end{figure}

From the microscopic colloid perspective, the swimmers form a kinetic boundary layer \cite{ForceBoundary2015} on the wall with directed motion as shown in Fig.~\ref{fig:Fig_BLnm}. More specifically, on a microscopic scale close to the wall, there is net polar order $\bmm=\int P\bq d\bq\neq 0$, even though the nematic orientation field has no polar order in the bulk. This boundary layer structure for two cases $\chi_R=0.4$ and $\chi_R=1.6$ are shown in Fig.~\ref{fig:Fig_BLnm}, with the FEM solution to the probability density $P(z,\theta)$.

\bibliography{ref}

\end{document}